# Influence length and space-time correlation between earthquakes

P. Tosi[1], V. De Rubeis[1], V. Loreto[2] and L. Pietronero[2]

*1 Istituto Nazionale di Geofisica e Vulcanologia (INGV), Via di Vigna Murata 606, 00143 Rome, Italy*

*2 "La Sapienza" University, Physics Department, and INFM, Center for Statistical Mechanics and Complexity, P.le A. Moro 5, 00185 Rome, Italy*

**Short and long range interactions between earthquakes are attracting increasing interest[1,2,3,4]. Scale invariant properties of seismicity[5,6,7,8] in time, space and energy argue for the presence of complex triggering mechanisms[9] where, like a cascade process, each event produces aftershocks[10]. A definitive method to assess any connection between two earthquakes separated in time and distance does not exist. Here we propose a novel method of data analysis that, based on the space-time combined generalization of the correlation integral[11] leads to a self-consistent visualization and analysis of both spatial and temporal correlations. When analyzing global seismicity we discovered a universal relation linking the spatial Influence Length $R_i$ of a given earthquake to the time $\tau$ elapsed from the event itself: $R_i \approx \tau^{-\alpha}$, with $\alpha \cong 0.55 \pm 0.05$. Following an event, time correlations (i.e. causality effects) exist in a region of radius $R_i$ that shrinks over time, suggesting a long-range dissipating stress transfer. A different process is acting in the short-range where events are randomly set, evidencing a sub-diffusive growth[12,13,14] of the seismogenic zone.**

Earthquakes appear to occur in clusters with scale invariant patterns in both space and time [7,8,15]. To date, analysis has tended to focus separately on either spatial or temporal correlations, with some notable exceptions [16,17]. Many earthquake properties point to a hierarchical organization, suggesting a connection among events that could be explained physically by stress transfer mechanisms and a scale invariant fracturing of the crust. Due to the complexity of the phenomenon, the influence of an earthquake over surrounding areas cannot be simply assessed, especially for long range effects. Seismicity is a multidimensional scale invariant process and every catalogue choice corresponds to an implicit multidimensional sizing (a defined window in energy, time and space: often dependent on data availability). In our opinion all these features are inextricably linked[18], calling for a combined analytical approach that can reveal important features previously not apparent. In this paper, a new method that leads to a self-consistent analysis and visualization of both spatial and temporal correlations is introduced. We have defined the *space-time combined correlation integral* as:

$$C(r,\tau) = \frac{2}{N(N-1)} \sum_{i=1}^{N-1} \sum_{j=i+1}^{N} \left( \Theta\left(r - \|\mathbf{x}_i - \mathbf{x}_j\|\right) \cdot \Theta\left(\tau - \|t_i - t_j\|\right) \right)$$

where $\Theta$ is the Heaviside step function ($\Theta(x) = 0$ if $x \leq 0$ and $\Theta(x) = 1$ if $x > 0$) and the sum counts all pairs whose spatial distance $\|\mathbf{x}_i - \mathbf{x}_j\| \leq r$ and whose time distance $\|t_i - t_j\| \leq \tau$ . When applied over all possible values of $\tau$ or $r$ , the well-known Grassberger-Procaccia correlation integral[11] is returned. Such a definition takes into account the distribution of all time intervals and epicentral inter-distances between all pairs of events, irrespective of the relationship between the main event and any aftershock.

From the space-time combined correlation integral we define the *time correlation dimension* for sets of events within space-time distances $r$ and $\tau$ , as:

$$D_t(r,\tau) = \frac{\partial \log C(r,\tau)}{\partial \log \tau} .$$

Similarly we define the *space correlation dimension*, for sets of events within space-time distances $r$ and $\tau$ , as:

$$D_s(r,\tau) = \frac{\partial \log C(r,\tau)}{\partial \log r} .$$

If $C(r,\tau)$ was a pure power-law in both variables, then $D_t$ and $D_s$ would correspond to the temporal and spatial fractal dimensions, respectively. More generally, the behaviour of $D_t$ and $D_s$ as a function of $r$ and $\tau$ will characterise the clustering features of earthquakes in space and in time. The possibility of handling a combined space-time correlation function allows for the investigation of seismic features where space and time are inextricably linked[9,19]. It is important to remark how the advantage of the method of analysis proposed lies in the possibility to simultaneously consider space and time correlations without *a priori* hypothesis: in fact, the method does not require any arbitrary choice of space or time windows.

We have started our analysis by considering world wide seismicity with the intention of making clear the most general space-time correlations. Subsequently, we compare our results with those obtained using local catalogues. From global seismic records we have selected epicenters in the time period between 1973 and 2002, with magnitudes greater than 5 (National Earthquake Information Center, U.S.G.S). This catalogue selection was conditioned by a criteria of completeness and it presents medium to high magnitude distribution; events mark plate boundaries at a global spatial scale.

Figure 1.a shows a colour-coded map of the *time correlation dimension* $D_t$ as a function of space and time. The colour coding of each pixel quantifies the time correlations existing between events occurring within a given distance and time interval. $D_t \cong 1$ corresponds to the random occurrence of events, while a lesser value of $D_t$ indicates time clustering. In Fig. 1.b the same key applies to the colour-coded map of the *space correlation dimension* $D_s$ as a function of space and time. In this case, values of $D_s$ approaching 2.0 on the scale identify sets of events uniformly distributed in space, while lower values of $D_s$ indicate a clustered spatial distribution of the epicenters (always within a given distance and time interval). It is remarkable to note how the patterns observed in Figs. 1.a and 1.b, which correspond to non-trivial structures in space and time for the correlations, significantly support our realisation that earthquakes do interact in space and in time. In order to check this hypothesis we have applied the same analysis to the global catalogue after reshuffling the time and epicenter locations of the events (keeping epicentral coordinates fixed and mixing occurrence dates). The results reported in Fig. 1.c and 1.d show that all patterns vanish, evidencing constant high values of $D_t$ and $D_s$ at all distance and time intervals. These results point strongly in favour of non-trivial interactions among earthquakes. Let us now comment on specific features emerging from our analysis.

**Time clustering**. In figure 1.a two main domains appear: one at shorter inter-distances with low $D_t$ (time clustering) representing a causal connection; the other with $D_t \cong 1$ indicating a random time occurrence of events. The boundary between these two domains is not sharp but, by defining the randomness limit at $D_t = 0.8$, a functional relationship can be extracted between the spatial separation $r$ and the temporal distance $\tau$. The results reported in Figure 2 indicate dependence of $r \propto \tau^{-\alpha}$, with $\alpha \cong 0.55 \pm 0.05$. If we now interpret $\tau$ as the time elapsed since a given earthquake and we define the distance from its epicenter $r$ as Influence Length $R_i$, we could argue that the functional relation $R_i \propto \tau^{-\alpha}$ describes how the size of the region causally connected to the given earthquake changes over time. In particular, a power-law shrinkage of the region can be seen. If compared to a homogeneous time distribution, this area of influence can be interpreted as a region of modified probability of earthquake time occurrence. Valid for worldwide seismicity, this relation has been verified for local catalogues with different magnitude ranges and tectonic settings (see captions of figures 2.a and 2.b for details). In Fig. 2.b it is important to note that raising the threshold level of lower magnitude data does not alter the power-law behaviour $R_i$, with approximately the same exponent and a pre-factor appearing dependent on the magnitude cut-off. In perspective, the relation seems to express a universal feature of the response properties of the earth crust to the occurrence of an earthquake.

**Space clustering** properties are shown in Figure 1.b for the worldwide seismicity catalogue: different domains are easily recognised. At short distances a high *space correlation dimension* behaviour is clearly separated from space clustering ($0 < D_s < 1$) that is present at greater distances: both conditions last for inter-time up to 100 days. There is no clear demarcation between the two regions but, when fixing the limit of clustering at $D_s = 1$, it appears that the area with high correlation dimension is evolving with time. In particular, the line separating randomly filled areas from those with space clustering follows the relation $\log r = 0.1 \log \tau + 1.2$ (with $\tau$ and $r$ defined as above). Localisation errors play an important role at short spatial ranges, giving high values of $D_s$, but the increasing limit detected advocates the presence of a physical process. The separation line defines a radius $R_o$, slowly growing in time, within which seismic events are spatially uniformly distributed. This finding is in agreement with the accepted migration of aftershocks away from a main shock[20] based on the interpretation of uniformly spatially distributed events as aftershocks. Many authors have described this migration in terms of a law $\overline{d}(t) \sim t^H$, where $\overline{d}(t)$ is the mean distance between main event and aftershocks occurring after time $t$, with an exponent $H < 0.5$ corresponding to a sub-diffusive process[12,13,14] often observed for local situations. Different behaviour can be observed over larger spatial ranges depending on the time intervals between events.

We summarize the results in Figure 3 where we recognize a 'near field' domain, evidenced by the behaviour of the space correlation dimension, and a 'far field' domain, defined by the behaviour of the time correlation dimension. Both domain ranges evolve in time: the first one slowly increasing and the latter quickly shrinking. An interpretation of this scenario in terms of stress transfer mechanisms[21] is possible.

A given earthquake induces different kinds of stress transfer mechanisms that can generally be categorised as either coseismic or postseismic. The first group is based on the elastic properties of the crust and can be either static or dynamic. Postseismic stress transfer (sometimes referred to as quasi-static) is associated with the slow viscous relaxation of the lower part of the crust and the upper part of the mantle. The debate is wide open as to which of these mechanisms is principally responsible for triggering an earthquake. It is generally accepted[4] that stress changes $d\sigma$ decay over distance $s$ as the power-law $d\sigma \propto |s|^{-a}$ with the exponent $a$ dependent on the specific mechanisms of stress transfer and on the lithosphere rheology. A general distinction can be made between 'near field' behaviour, occurring at distances from the triggering event in the order of the fault length (and more generally of the order of the size of the seismogenic structure), and 'far field' or long-range behaviour. Many factors drive stress transfer and seismicity in the near field such as: complicated source mechanisms, pre-existing weakness zones, heterogeneity of the fault plane[22] and fluid migration[23]. When focusing on the far field, i.e. on what happens in a region much larger than the fault length, many of these factors integrate out and a general statistical description appears possible. This longer range indicates an ever changing stress field that tends to weaken over time, either by earthquake occurrences or by aseismic creep (slow slips not generating elastic waves) and, more generally, by all the mechanisms falling under the denomination of stress leakage[24, 25]. Depending on the distance from the main event, an alteration in stress can statistically affect the failure probability[26] and, hence, the seismicity rate[27]. Greater the stress change detected, greater should be the seismicity rate change. Consequently, for very small stress changes the triggering/shadowing effect is negligible and suggests the existence of some sort of elastic threshold, below which the stress change should not be able to affect the seismicity rate. Such a lower cut-off could be identified, for instance, with the level of 'tidal stress' that is induced by the distortion of the earth caused by the pull of the sun and moon. Typical values of tidal stress changes are in the order of 0.01 bars, and do not directly influence seismicity[28]. Adoption of this threshold allows identification of a length scale, $R(t)$, defined as the distance from the main event's epicenter for which the stress change falls below the lower cut-off. We interpret this length scale $R(t)$ as the radius of the region causally connected to the main event $t$ seconds after the event itself. Since the level of stress change drops over time one should expect that $R(t)$ decreases over time. This is what can be observed if we interpret $R(t)$ and $t$ as the previously defined $R_i$ and $\tau$.

In summary, we have introduced a new statistical tool, the combined space-time correlation integral, which allows us to perform a simultaneous and self-consistent investigation of the spatial and temporal correlation properties of earthquakes. This tool leads to the discovery, visualization and deep analysis of the complex interrelationships existing between the spatial distribution of epicenters and their occurrence in time. Three main results emerged: The comparison between space and time correlations performed on the worldwide seismicity catalogue and the corresponding reshuffled catalogue, strongly suggests that earthquakes do interact spatially and temporally. From the study of time clustering, a new universal relation linking the so-called 'Influence Length' of an earthquake and the time elapsed since its occurrence is discovered. Finally, analysis of the space clustering reveals the existence of a region where events are uniformly distributed in space. The size of this region increases slowly with time, supporting existing theories on aftershock diffusion. Together the results set the basis for basis further validation on both worldwide and local scales, as well as for suitable modelling[29]. Beyond relevance in seismology it is worth stressing how our contribution could be potentially important in a wider context, where understanding the interplay of spatial and temporal correlations is crucial for the correct interpretation of phenomena such as solar flares, acoustic emissions, dynamical systems theory, and the dynamics of extended systems in physics and biology.

**Acknowledgements** We thank M. Cocco and A. Rovelli for comments and useful discussion.

**Correspondence** and requests for materials should be addressed to P.T. (tosi@ingv.it)

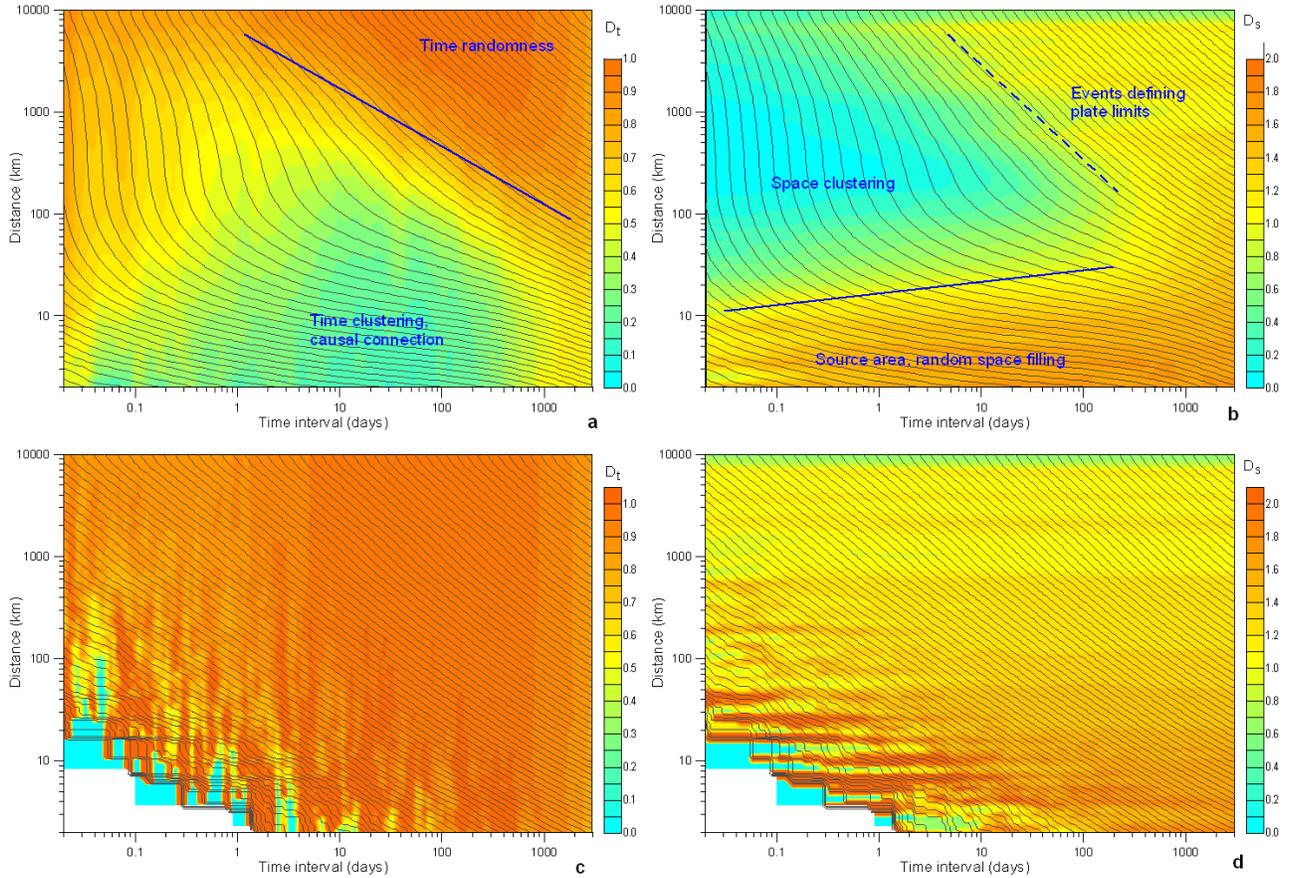

**Figure 1** Correlations in global seismicity. **a** = Temporal correlations of global seismicity. Values of space-time combined correlation integral are represented by dark contour lines; horizontal direction derivative of this surface is the *time correlation dimension* (coloured shaded contour). $D_t$ is low – thus indicating correlation – in spatial ranges which clearly depend on time. The spatial range correlation tends to increase during the time interval from within hours to a few days after each event, although without a marked limit. This period of time is characterised by a perturbation propagation ranging from the source neighbourhood to hundreds of kilometres away. The time correlation spatial range decreases during the interval from few days to circa $10^4$ days, clearly defining a region of seismic time related activity from random event occurrence. (Figure 2 shows this correlation boundary with further details.) **b** = Spatial correlation of global seismicity. The epicenter *space correlation dimension* ranges from extreme clustering ($D_s = 0$) to random space distribution ($D_s = 2$). At short distances in the order of seismic source dimensions (10-30 km), there are prevailing high values ($D_s > 1.5$): this can be interpreted as the elevated random space filling tendency of epicenters over the seismogenic zone. Separation between space randomness and space clustering as a function of time is evidenced (continuous line). The upper limit (dotted line) shows a decreasing range in time of around two orders of magnitude. This can be interpreted as a space connection: seismic sources act like a seismic attractor over a space whose range is shrinking over time. It should be noted that low space correlation does not imply the uniqueness of a seismic source, but that seismicity is allowed on specific correlated structures. The disappearance of clustering with time leaves room to a general $D_s = 1$ correlation dimension, interpreted as the activity of seismicity on plate boundaries. **c,** Temporal correlation and **d,** Spatial correlation of the global catalogue after a reshuffling procedure. The patterns observed in Figs. 1.a and 1.b completely disappear evidencing random distributions both in space and in time. This result confirms the presence of a physical process of interaction between earthquakes, linking the space and time of their occurrence.

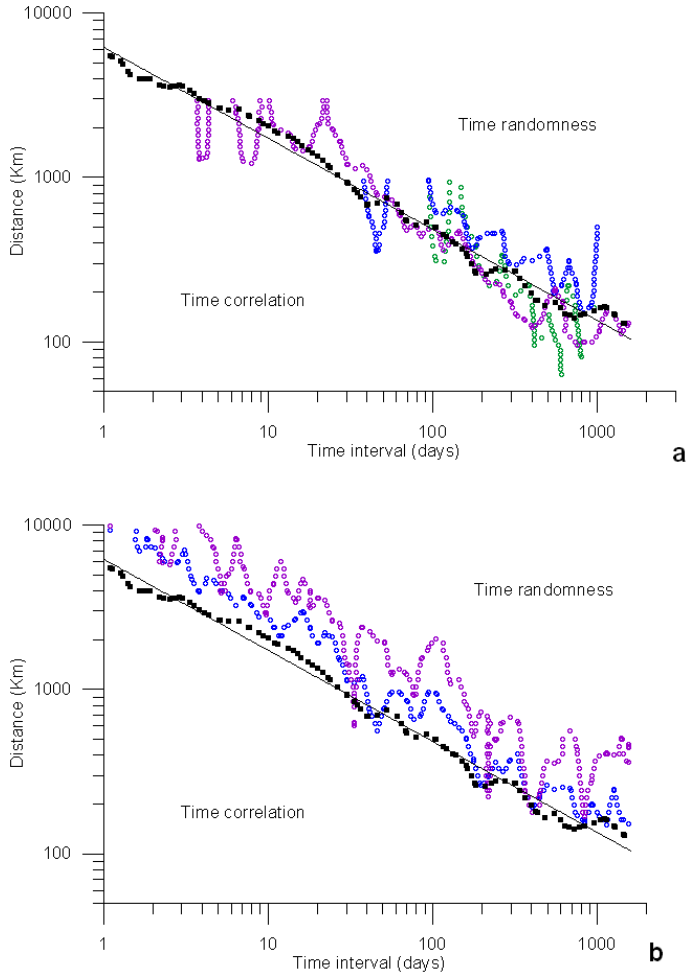

**Figure 2** Time correlation and Influence Length. **a,** The limit separating time correlation from time randomness (arbitrarily fixed to $D_t = 0.8$) as a function of distance is shown. It follows the law (continuous line) $\log r = -0.55 \log \tau + 3.8$ where $\tau$ is expressed in days and $r$ in km, in relation to the world wide seismic catalogue (black dots, years from 1973 to 2002, $m_b \geq 5$, National Earthquake Information Center, U.S.G.S). This relation places a strong constraint on time relations among events, evidencing how distance plays a dynamical role. In particular, the relation can be read as defining an Influence Length shrinking over time with a power-law behaviour. For relatively short spatial ranges (around 100 km) events are time clustered and correlated for long time intervals (around 3 years). Over longer distances time correlation lasts for a short period (less than 30 days for 1000 km). $D_t = 0.8$ points for other catalogues are represented with different colours. Green: California (1980 to 2002, $M_d \geq 3$, Berkeley Seismological Laboratory). Blue: Italy (1983-2002, $M_l \geq 3$, Istituto Nazionale di Geofisica e Vulcanologia). Violet: Tibet Region (1974-2003; $m_b \geq 4.5$, Advanced National Seismic System). **b,** $D_t = 0.8$ values for the global catalogue at increasing magnitude limits. Black: $m_b \geq 5$. Blue: $m_b \geq 5.5$. Violet: $m_b \geq 6$. Regardless of differing regions, tectonic settings and magnitude limits, all catalogues show a similar behaviour with a magnitude dependent prefactor.

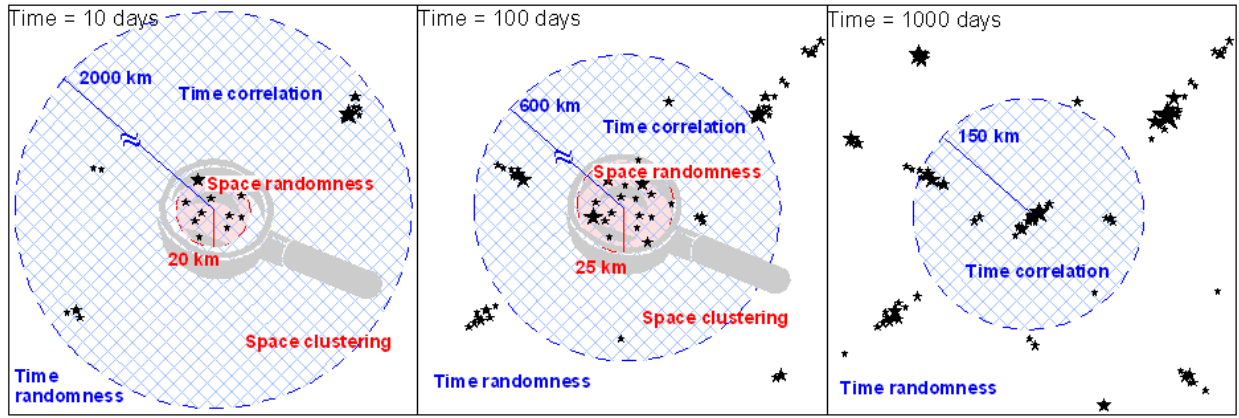

**Figure 3** Dynamical evolution. At time t=0 an earthquake occurs in the centre of the represented area. Within one day the stress perturbation reaches long distances (typically of the order of thousands of kilometres for medium-large magnitudes) and seismicity is modified inside a range that we name Influence Length $R_i$ (blue circle). This causal connection region then shrinks over time with a power-law behaviour. Meanwhile, in the near-field, aftershocks slowly diffuse in random patterns, within an area of increasing size (red circle).